\documentclass{llncs}
\usepackage{amsmath}

\usepackage{courier}            
\usepackage[scaled]{helvet} 
\usepackage{xspace}
\usepackage{proof}
\usepackage{graphicx}
\usepackage{amsmath}
\usepackage{amscd}
\usepackage{amssymb}
\usepackage{float}
\usepackage{epsfig}
\usepackage{xcolor}
\usepackage{color}
\usepackage{appendix}
\usepackage{hyperref}
\usepackage{adjustbox}
\usepackage{subcaption}

\usepackage{multirow}
\usepackage{wrapfig}

\usepackage[T1]{fontenc}
\usepackage[kerning,spacing]{microtype}
 
 \usepackage{tikz}
 
\usepackage{tikzscale}
\makeatother
\usetikzlibrary{shapes,arrows}
\tikzstyle{decision} = [diamond, draw, fill=blue!20, 
text width=4.5em, text badly centered, node distance=3cm, inner sep=0pt]
\tikzstyle{block} = [rectangle, draw, fill=white!20, node distance = 6cm, text centered, rounded corners, minimum height=2em]
\tikzstyle{line} = [draw, -latex']
\tikzstyle{cloud} = [draw, ellipse,fill=red!20, node distance=5cm,
minimum height=2em]

\definecolor{redcolor}{rgb}{0.8,0,0}
\newcommand{\TODO}[1]{{\color{redcolor} [TODO: \textsl{#1}]}}

\newcommand{\hide}[1]{} 
\newcommand{\C}[1]{\lstinline!#1!}

\newcommand{\Sk}{\textsc{Sketch}}
\newcommand{\lang}{\textsc{SyntRec}}

\newcommand{\seclabel}[1]{\label{sec:#1}}

\newcommand{\longsecref}[1]{Section~\ref{sec:#1}}

\newcommand{\figlabel}[1]{\label{fig:#1}}
\newcommand{\longfigref}[1]{Figure~\ref{fig:#1}}

\newcommand{\secref}{\longsecref}
\newcommand{\figref}{\longfigref}

\numberwithin{equation}{section}

  \newenvironment{formula}
 {\small \[}
 {\]}
 \newenvironment{formulaarr}
 {\small \vspace{-7pt} \begin{eqnarray*}}
 {\end{eqnarray*}}

\usepackage[T1]{fontenc}

\definecolor{dkgreen}{rgb}{0,0.3,0}
\definecolor{gray}{rgb}{0.5,0.5,0.5}
\definecolor{mauve}{rgb}{0.58,0,0.82}
\definecolor{light-gray}{gray}{0.80}

\usepackage{listings}
\lstdefinelanguage{sketch}{
alsoletter={?},
  morekeywords = {
       bool, harness, data, int, bit, bool, adt, new, return, assert , assume, let, case, switch, choose, in, repeat\_case, generator, cons?, fields?, case?, ??},
 commentstyle=\itshape,
  morecomment=[s]{/*}{*/},
}

\lstdefinestyle{nonumstyle}{
  language=sketch,
  columns=flexible,
  basicstyle=\fontfamily{lmss}\selectfont\footnotesize,
  numbers=none,
  numbersep=3pt,
  numberstyle=\tiny,
  stepnumber=1,
  tabsize=2,
  breaklines=false,
  breakatwhitespace=true,
  commentstyle=\color{cyan},
  mathescape=true,
  escapeinside={(*}{*)}
}
\lstdefinestyle{numstyle}{
  language=sketch,
  columns=flexible,
  basicstyle=\fontfamily{lmss}\selectfont\footnotesize,
  numbers=left,
  numbersep=5pt,
  numberstyle=\tiny,
  stepnumber=1,
  tabsize=2,
  breaklines=false,
  breakatwhitespace=true,
  commentstyle=\color{cyan},
  mathescape=true,
  escapeinside={(*}{*)}
}

\renewcommand{\scriptsize}{\fontsize{8.5}{9}\selectfont}
\makeatletter
\lst@AddToHook{TextStyle}{\let\lst@basicstyle\scriptsize\fontfamily{lmss}\selectfont}
\makeatother

\usepackage{url}

\usepackage{rotating}

\usepackage{enumitem}

\newcommand{\undefined}{{\textit{ndef}}}

\begin{document}

\title{Synthesis of Recursive ADT Transformations from Reusable Templates}
\titlerunning{Synthesis of Recursive ADT Transformations}  
%
\author{Jeevana Priya Inala\inst{1} \and Nadia Polikarpova\inst{1} \and Xiaokang Qiu\inst{2} \and
Benjamin S. Lerner\inst{3} \and Armando Solar-Lezama\inst{1}}
\authorrunning{Jeevana P. Inala et al.} 
%

%
\institute{MIT~~
\email{\{jinala,  polikarn, asolar\}@csail.mit.edu}
\and
Purdue University~~
\email{xkqiu@purdue.edu}
\and
Northeastern University~~
\email{blerner@ccs.neu.edu}
}

\maketitle              

\begin{abstract}
Recent work has proposed a promising approach to improving scalability of program synthesis
by allowing the user to supply a syntactic template 
that constrains the space of potential programs.
Unfortunately, creating templates often requires nontrivial effort from the user,
which impedes the usability of the synthesizer.
We present a solution to this problem in the context of recursive transformations on algebraic data-types.
Our approach relies on \emph{polymorphic synthesis constructs}:
a small but powerful extension to the language of syntactic templates,
which makes it possible to define a program space in a concise and highly reusable manner,
while at the same time retains the scalability benefits of conventional templates.
This approach enables end-users to reuse predefined templates from a library
for a wide variety of problems with little effort.
The paper also describes a novel optimization that further improves the performance and the scalability of the system. 
We evaluated the approach on a set of benchmarks that most notably includes desugaring functions for lambda calculus, 
which force the synthesizer to discover Church encodings for pairs and boolean operations.

\end{abstract}



\newcommand{\mc}[1]{\text{\C{#1}}}
\newcommand{\rcase}{\mc{repeat_case}}
\newcommand{\optname}{\emph{inductive decomposition}}
\newcommand{\tdc}{\emph{PSC}}
\section{Introduction}\seclabel{intro}

Recent years have seen remarkable advances in tools and techniques for automated synthesis of recursive programs~\cite{KneussKKS13,AlbarghouthiGK13,OseraZ15,FeserCD15,synquid}.
These tools take as input some form of \emph{correctness specification}
that describes the intended program behavior,
and a set of building blocks (or \emph{components}).
The synthesizer then performs a search in the space of all programs that can be built from the given components
until it finds one that satisfies the specification.
The biggest obstacle to practical program synthesis is that this search space grows extremely fast with the size of the program
and the number of available components.
As a result, these tools have been able to tackle only relatively simple tasks,
such as textbook data structure manipulations. 

Syntax-guided synthesis (SyGuS)~\cite{sygus} has emerged as a promising way to address this problem.
SyGuS tools, such as \Sk{}~\cite{sketchthesis} and Rosette~\cite{TorlakB13,TorlakB14} leverage a user-provided syntactic \emph{template} 
to restrict the space of programs the synthesizer has to consider,
which improves scalability and allows SyGus tools to tackle much harder problems.
However, the requirement to provide a template for every synthesis task significantly impacts usability.

This paper shows that, at least in the context of recursive transformations on algebraic data-types (ADTs), it is possible to get the best of both worlds.
Our first contribution is a new approach to making syntactic templates highly reusable by relying on \emph{polymorphic synthesis constructs} (\tdc{}s).
With \tdc{}s, a user does not have to write a custom template for every synthesis problem, 
but can instead rely on a generic template from a library.
Even when the user does write a custom template, the new constructs make this task simpler and less error-prone.
We show in \secref{experiments} that all our 23 diverse benchmarks are synthesized using just 4 different generic templates from the library. 
Moreover, thanks to a carefully designed type-directed expansion mechanism,
our generic templates provide the same performance benefits during synthesis as conventional, program-specific templates. 
Our second contribution is a new optimization called \emph{inductive decomposition}, 
which achieves asymptotic improvements in synthesis times for large and non-trivial ADT transformations.
This optimization, together with the user guidance in the form of reusable templates,
allows our system to attack problems that are out of scope for existing synthesizers.


We implemented these ideas in a tool called \lang{}, 
which is built on top of the open source \Sk{} synthesis platform~\cite{sketchHome}. 
Our tool supports expressive correctness specifications that can use arbitrary functions to constrain the behavior of ADT transformations. 
Like other expressive synthesizers, such as \Sk{}~\cite{sketchthesis} and Rosette~\cite{TorlakB14,TorlakB13},
our system relies on exhaustive bounded checking to establish whether a program candidate matches the specification. 
While this does not provide correctness guarantees beyond a bounded set of inputs, 
it works well in practice and allows us to tackle complex problems, 
for which full correctness is undecidable and is beyond the state of the art in automatic verification.
For example, our benchmarks include desugaring functions from an abstract syntax tree (AST) into a simpler AST,
where correctness is defined in terms of interpreters for the two ASTs.
As a result, our synthesizer is able to discover Church encodings for pairs and booleans,
given nothing but an interpreter for  the lambda calculus.
In another benchmark, we show that the system is powerful enough to synthesize a type constraint generator for a simple programming language 
given the semantics of type constraints. 
Additionally, several of our benchmarks come from transformation passes implemented in our own compiler and synthesizer.

\section{Overview}\seclabel{overview}
In this section, we use the problem of desugaring a simple language to illustrate the main features of \lang{}. 
Specifically, the goal is to synthesize a function \C{dstAST desugar(srcAST src)\{ $\ldots$ \}},
which translates an expression in source AST into a semantically equivalent expression in  destination AST. 
Data type definitions for the two ASTs are shown in \figref{runningex}: 
the type \C{srcAST} has five \emph{variants} (two of which are recursive),
while \C{dstAST} has only three.
In particular, the source language construct \C{BetweenS(a, b, c)}, which denotes \C{a < b < c}, 
has to be desugared into a conjunction of two inequalities.
Like case classes in Scala, data type variants in \lang{} have named fields. 


\vspace{-5pt}
\begin{figure}[t]
\footnotesize
\begin{adjustbox}{max width=\textwidth}
\begin{tabular}{lcl}
\footnotesize
\begin{lstlisting}
  adt srcAST{ 
    NumS{ int v; }
    TrueS{ }
    FalseS{ }
    BinaryS{ opcode op; srcAST a; srcAST b;}
    BetweenS{ srcAST a; srcAST b; srcAST c;}}
\end{lstlisting}
&~~&
\footnotesize
\begin{lstlisting}
  adt dstAST{
    NumD{ int v; }
    BoolD{ bit v; }
    BinaryD{ opcode op; dstAST a; dstAST b;}}
    
  adt opcode{ AndOp{} OrOp{} LtOp{}}
\end{lstlisting}
\end{tabular}
\end{adjustbox}
\caption{ADTs for two small expression languages}
\vspace{-15pt}
\figlabel{runningex}
\end{figure}

\paragraph{\textbf{Specification.}} 
The first piece of user input required by the synthesizer is the specification of the program's intended behavior.
In the case of \C{desugar}, we would like to specify that the desugared AST is semantically equivalent to the original AST,
which can be expressed in \lang{} using the following constraint:
\vspace{-5pt}
\begin{center}
\begin{tabular}{c}
\begin{lstlisting}
assert( srcInterpret(exp) == dstInterpret(desugar(exp)) )
\end{lstlisting}
\end{tabular}
\end{center}
\vspace{-5pt}
This constraint states that interpreting an arbitrary source-language expression \C{exp} (bounded to some depth) must be equivalent to desugaring \C{exp} and interpreting the resulting expression in the destination language. 
Here, \C{srcInterpret} and \C{dstInterpret} are regular functions written in \lang{} and defined recursively over the structure of the respective ASTs in a straightforward manner.
As we explain in \secref{synthesis}, our synthesizer contains a novel optimization called \optname{} that can take advantage of the structure of the above specification to significantly improve the scalability of the synthesis process. 

\paragraph{\textbf{Templates.}}
The second piece of user input required by our system is a syntactic \emph{template}, which describes the space of possible implementations.
The template is intended to specify the high-level structure of the program, leaving low-level details for the system to figure out.
In that respect, \lang{} follows the SyGuS paradigm~\cite{sygus};
however, template languages used in existing SyGuS tools, such as \Sk{} or Rosette,
work poorly in the context of recursive ADT transformations.

For example, \figref{sketchtemp} shows a template for \C{desugar} written in \Sk{},
the predecessor of \lang{}. It is useful to understand this template as we will show, later, how the new language features in \lang{} allow us to write the same template in a concise and reusable manner.
This template uses three kinds of \emph{synthesis constructs} already existing in \Sk{}:
a \emph{choice} (\C{choose($e_1$,...,$e_n$)}) must be replaced with one of the expressions $e_1,\ldots,e_n$;
a \emph{hole} (\C{??}) must be replaced with an integer or  a boolean constant;
finally, a \emph{generator} (such as \C{rcons}) can be thought of as a macro, which is inlined on use,
allowing the synthesizer to make different choices for every invocation%
\footnote{Recursive generators, such as \C{rcons}, are unrolled up to a fixed depth, which is a parameter to our system.}.
The task of the synthesizer is to fill in every choice and hole in such a way that the resulting program satisfies the specification. 

The template in \figref{sketchtemp} expresses the intuition 
that \C{desugar} should recursively traverse its input, \C{src}, replacing each node with some subtree from the destination language.
These destination subtrees are created by calling the recursive, higher-order generator \C{rcons} (for ``recursive constructor'').
\C{rcons(e)} constructs a nondeterministically chosen variant of \C{dstAST},
whose fields, depending on their type, are obtained
either by recursively invoking \C{rcons},
by invoking \C{e} (which is itself a generator),
or by picking an integer or boolean constant.
For example, one possible instantiation of the template \C{rcons(choose(x, y, src.op))} \footnote{When an expression is passed as an argument to a higher-order function that expects a function parameter such as \C{rcons}, it is automatically casted to a \emph{generator lambda} function. Hence, the expression will only be evaluated when the higher-order function calls the function parameter and each call can result in a different evaluation.}
can lead to  \C{new BinaryD(op = src.op, a = x, b = new NumD(5))}.
Note that the template for \C{desugar} provides no insight on how to actually encode each node of \C{scrAST} in terms of \C{dstAST},
which is left for the synthesizer to figure out.
Despite containing so little information, the template is very verbose: in fact, more verbose than the full implementation!
More importantly, this template cannot be reused for other synthesis problems,
since it is specific to the variants and fields of the two data types.
Expressing such a template in Rosette will be similarly verbose.

\begin{figure}[t]
\begin{adjustbox}{max width = \textwidth}
\begin{tabular}[T]{lcr}
\begin{lstlisting}
dstAST desugar(srcAST src){
  switch(src) {
  case NumS: 
    return rcons(src.v);
  ... /* Some cases are elided */
  case BinaryS:
   	dstAST a = desugar(src.a), b = desugar(src.b); 
    return rcons(choose(a, b, src.op));
  case BetweenS:
   	dstAST a = desugar(src.a), b = desugar(src.b), 
    				c = desugar(src.c);
    return rcons(choose(a, b, c));
}}
\end{lstlisting}
& ~~ &
\begin{lstlisting}
generator dstAST rcons(fun e) {
  if (??) return e();
  if (??) {
    int val = choose(e(), ??);
   	return new NumD(v = val); }
  if (??) {
   	bit val = choose(e(), ??);
    return new BoolD(v = val);}
  if (??) {
   	dstAST a = rcons(e);
    dstAST b = rcons(e);
    opcode op = choose(e(), new AndOp(),...,
   						 new LtOp());
   	return new BinaryD(op = op, a= a, b = b);}
}
\end{lstlisting}
\end{tabular}
\end{adjustbox}
\caption{Template for \C{desugar} in \Sk{}}
\vspace{-15pt}
\figlabel{sketchtemp}
\end{figure}

\paragraph{\textbf{Reusable Templates.}} \lang{} addresses this problem by extending the template language with \emph{polymorphic synthesis constructs} (\tdc{}s),
which essentially support parametrizing templates by the structure of data types they manipulate.
As a result, in \lang{} the end user can express the template for \C{desugar} with a single line of code:
\vspace{-5pt}
\begin{lstlisting}
dstAST desugar(srcAST src) { return recursiveReplacer(src, desugar);  }
\end{lstlisting}
\vspace{-5pt}
Here, \C{recursiveReplacer} is a reusable generator defined in a library;
its code is shown in \figref{tmp-sol}.
When the user invokes \C{recursiveReplacer(src,desugar)},
the body of the generator is specialized to the surrounding context,
resulting in a template very similar to the one in \figref{sketchtemp}.
Unlike the template in \figref{sketchtemp}, however, \C{recursiveReplacer} is not specific to \C{srcAST} and \C{dstAST},
and can be reused with no modifications
to synthesize desugaring functions for other languages,
and even more general recursive ADT transformations.
Crucially, even though the reusable template is much more concise than the \Sk{} template,
it does not increase the size of the search space that the synthesizer has to consider,
since all the additional choices are resolved during type inference.
\figref{tmp-sol} also shows a compacted version of the solution for \C{desugar}, 
which \lang{} synthesizes in about 8s.
The rest of the section gives an overview of the \tdc{}s used in \figref{tmp-sol}.

\newcommand{\caseh}{\textbf{case?}}
\newcommand{\fieldsh}{\textbf{fields?}}
\newcommand{\consh}{\textbf{cons?}}
\newcommand{\casehc}{\C{case?}}
\newcommand{\fieldshc}{\C{fields?}}
\newcommand{\conshc}{\C{cons?}}

\lstset{style=numstyle}

\begin{figure}[t]
\begin{adjustbox}{max width=\textwidth}
\begin{tabular}[T]{rcr}
\begin{lstlisting}
generator T recursiveReplacer<T, Q>(Q src,
							fun rec) {
   switch(src){
     (*\caseh{}*):(*\label{line:case}*)
         T[ ]  a = map(src.(*\fieldsh{}*), rec);(*\label{line:fields}*)
        	return rcons(choose(a[??],
				        	 field(src)));
}}}
generator T rcons<T>(fun e) {
	if (??) return e();
	else return new (*\consh{}*)(rcons(e)); (*\label{line:cons}*)
}
generator T field<T,S>(S e) {
	return (e.(*\fieldsh{}*)) [??];
}
\end{lstlisting}
&~~~~~~~~~&
\begin{lstlisting}[gobble=2]
  dstAST desugar(srcAST src) {
    switch(src) {
      case NumS: return new NumD(v = src.v);
 	 	 case TrueS: return new BoolD(v = 1);
      case FalseS: return new BoolD(v = 0);
      case BinaryS:
        dstAST[2] a = {desugar(src.a), desugar(src.b)};
        return new BinaryD(op = src.op, a = a[1],
									         b = a[2]);
      case BetweenS:
        dstAST[3] a = {desugar(src.a), desugar(src.b),
        					desugar(src.c)};
        return new BinaryD(op = new AndOp(), 
          a = new BinaryD(op = new LtOp(), a = a[0],
								           b = a[1])
          b = new BinaryD(op = new LtOp(), a = a[1], 
									         b = a[2]));
  }}
\end{lstlisting}
\end{tabular}
\end{adjustbox}
\caption{Left: Generic template for \C{recursiveReplacer} Right: Solution to the running example}
\vspace{-15pt}
\figlabel{tmp-sol}
\end{figure}

\lstset{style=nonumstyle}

\paragraph{\textbf{Polymorphic Synthesis Constructs.}} 
Just like a regular synthesis construct, a \tdc{} represents a set of potential programs, 
but the exact set depends on the context
and is determined by the types of the arguments to a \tdc{} and its expected return type.
\lang{} introduces four kinds of \tdc{}s.


\noindent
1. A \textbf{\textit{Polymorphic Generator}} is a polymorphic version of a \Sk{} generator.
For example, \C{recursiveReplacer} is a polymorphic generator, parametrized by types \C{T} and \C{Q}.
When the user invokes \C{recursiveReplacer(src,desugar)},
\C{T} and \C{Q} are instantiated with \C{dstAST} and \C{srcAST}, respectively.

\noindent
2. \textbf{\textit{Flexible Pattern Matching}} (\C{switch(x) } \casehc{}\C{: e}) expands into pattern matching code specialized for the type of $x$. 
In our example, once \C{Q} in \C{recursiveReplacer} is instantiated with \C{srcAST},
the \casehc{} construct in Line \ref{line:case} expands into five cases (\C{case NumS}, ..., \C{case BetweenS})
with the body of \casehc{} duplicated inside each of these cases.


\noindent
3. \textbf{\textit{Field List}} (\C{e.}\fieldshc{}) expands into an array of all fields of type \C{$\tau$}
in a particular variant of \C{e}, 
where $\tau$ is derived from the context. 
Going back to \figref{tmp-sol}, Line \ref{line:fields} inside \C{recursiveReplacer} maps a function \C{rec} over a field list \C{src.}\fieldshc{};
in our example, \C{rec} is instantiated with \C{desugar},
which takes an input of type \C{srcAST}.
Hence, \lang{} determines that \C{src}.\fieldshc{} in this case denotes all fields of type \C{srcAST}.
Note that this construct is expanded differently in each of the five cases
that resulted from the expansion of \casehc{}.
For example, inside \C{case NumS}, this construct expands into an empty array (\C{NumS} has no fields of type \C{srcAST}),
while inside \C{case BetweenS}, it expands into the array \C{\{src.a, src.b, src.c\}}. 


\noindent
4. \textbf{\textit{Unknown Constructor}} (\C{new } \conshc{}\C{($e_{1}$, ..., $e_{n}$)}) expands into a constructor for some variant of type $\tau$, where $\tau$ is derived from the context, and uses the expressions $e_1, \ldots, e_n$ as the fields. 
In our example, the auxiliary generator \C{rcons} uses an unknown constructor in Line \ref{line:cons}. 
When \C{rcons} is invoked in a context that expects an expression of type \C{dstAST}, 
this unknown constructor expands into \C{choose(new NumD(...), new BoolD(...), new BinaryD(...))}. 
If instead \C{rcons} is expected to return an expression of type \C{opcode}, 
then the unknown constructor expands into \C{choose(new AndOp(),...,new LtOp())}. 
If the expected type is an integer or a boolean, this construct expands into a regular \Sk{} hole (\C{??}). 



Even though the language provides only four \tdc{}s, they can be combined in novel ways to create richer polymorphic constructs that can be used as library components.
The generators \C{field} and \C{rcons} in \figref{tmp-sol} are two such components.

\begin{wrapfigure}{r}{0.5\textwidth}
	\vspace{-0.8cm}
	\begin{adjustbox}{max width = 0.5\textwidth}
		$\begin{array}{rcl}
		P & := & \{ adt_{i} \}_i ~~ \{ f_{i} \}_i \\
		adt & := & \mc{adt}~name\left \{~~variant_{1}\ldots  variant_{n}~\right \} \\
		variant & := & name\left \{l_{1}:\tau_{1}\ldots ~l_{n}:\tau_{n}\right \} \\
		\theta & := & \tau ~~|~~ T~~|~~\theta[~] ~~|~~ fun~~| ~~\theta_1 \rightarrow \theta_2\\
		\tau & := & prim~~|~~name~~|~ \left \{l_{i}:\tau_{i}\right \}_{i<n}\\
		& & |~~\sum~name_{i}\left \{l_{k}^{i}:\tau _{k}^{i}\right \}_{k<n_i}\\
		prim & := & \mc{bit}~~|~~\mc{int}\\
		f & :=& ~~\overline{f}~~|~~\hat{f}~~|~~\hat{\hat{f}}\\
		\overline{f} & := & \tau_{out}~name\left(\{x_{i}:\tau_{i}\}_i\right)~~e~ \\
		\hat{f} & := & \mc{generator} ~\tau_{out}~name\left(\{x_{i}:\tau_{i}\}_i\right)~e~ \\
		\hat{\hat{f}} & := & \mc{generator}~~\theta_{out}~name\langle\{T_i\}_i\rangle\left(\{x_{i}:\theta_{i}\}_i \right)~~e~ \\
		e & := & ~~\overline{e}~~|~~\hat{e}~~|~~\hat{\hat{e}}\\
		\overline{e}&:=&~~x~~|~~\mc{let}~x:\theta =e_{1}~\mc{in}~~e_{2} ~~ | ~~ f(e)\\
		& & |~~\mc{switch} \left(x\right)\left \{~\mc{case}~~name_{i}:e_{i}~\right \}_i \\
		& & |~~e.l~~|~~\mc{new}~name(\{l_{i}=e_{i}\}_i) \\
		&& | ~~ \{\{e_i\}_i\} ~~|~~e_1[e_2] ~~|~~\mc{assert}(e)\\
		\hat{e} & := & ~~??~~|~~ \mc{choose}(\{e_i\}_i) | ~~\hat{f}(e)\\
		\hat{\hat{e}} &:=&~~\hat{\hat{f}}(e)~~|~~ \mc{new}~\mc{cons?}(\{e_i\}_i) ~~\\
		& & |~~ e.\mc{fields?}{} ~~|~~ \mc{switch}(x)\{ \mc{case?}{}: e \}\\
		\end{array}$
	\end{adjustbox}
	\caption{Kernel language}
	\vspace{-10pt}
	\figlabel{lang}
	\vspace{-0.3cm}
\end{wrapfigure}
The \C{field} component expands into an arbitrary field of type $\tau$, where $\tau$ is derived from the context.
Its implementation uses the \emph{field list} \tdc{} to obtain the array of all fields of type $\tau$, 
and then accesses a random element in this array using an integer hole.
For example, if \C{field(e)} is used in a context where the type of \C{e} is \C{BetweenS} and the expected type is \C{srcAST}, 
then \C{field(e)} expands into \C{\{e.a, e.b, e.c\}[??]} which is semantically equivalent to \C{choose(e.a, e.b, e.c)}.


The \C{rcons} component is a polymorphic version of the recursive constructor for \C{dstAST} in \figref{sketchtemp},
and can produce ADT trees of any type up to a certain depth.
Note that since \C{rcons} is a polymorphic generator, 
each call to \C{rcons} in the argument to the unknown constructor (Line \ref{line:cons}) 
is specialized based on the type required by that constructor and can make different non-deterministic choices. 
%
Similarly, it is possible to create other generic constructs such as iterators over arbitrary data structures. 
Components such as these are expected to be provided by expert users, while end users treat them in the same way as the built-in \tdc{}s. 
The next section gives a formal account of the  \lang{}'s language and the synthesis approach.

\section{\lang{} Formally}
\subsection{Language}\seclabel{language}
\figref{lang} shows a simple kernel language that captures the relevant features of \lang{}.  In this language, a program consists of a set of ADT declarations followed by a set of function declarations. 
The language distinguishes between a standard function  $\overline{f}$, a generator  $\hat{f}$ and a \emph{polymorphic generator} $\hat{\hat{f}}$.
Functions can be passed as parameters to other functions, but they are not entirely first-class citizens because they cannot be assigned to variables or returned from functions. Function parameters lack type annotations and are declared as type $fun$, but their types can be deduced from inference.
Similarly, expressions are divided into standard expressions that does not contain any unknown choices ($\overline{e}$), existing synthesis constructs in \Sk{} ($\hat{e}$), and the new \tdc{}s ($\hat{\hat{e}}$).  The language also has support for arrays with expressions for array creation ($\{e_1, e_2, ..., e_n\}$) and array access  ($e_1[e_2]$).  An array type is represented as $\theta[~]$.
In this formalism, we use the Greek letter  $\tau $  to refer to a fully concrete type and  $\theta $  to refer to a type that may involve type variables. The distinction between the two is important because \tdc{}s   can only be expanded when the types of their context are  known. We formalize ADTs as tagged unions $\tau = \sum variant_i$, where each of the variants is a record type $variant_i = name_{i}\left \{l_{k}^{i}:\tau _{k}^{i}\right \}_{k<n_i}.$
Note that ADTs in \lang{} are not polymorphic. The notation $\{a_i\}_i$ is used to denote the  $\{a_1, a_2,...\}$.

\subsection{Synthesis Approach}
Given a user-written program $\hat{\hat{P}}$ that can potentially contain \tdc{}s, choices and holes,  and a specification, the synthesis problem is to find a program $\overline{P}$ in the language that only contains standard expressions ($\overline{e}$) and  functions ($\overline{f}$). \lang{} solves this problem using a two step approach as shown below.:

\vspace{0.2cm}
\begin{adjustbox}{max width = \textwidth}
\begin{tikzpicture}[node distance = 3cm, auto]
\node [text centered, text width=5.5em, node distance=3cm] (p1) {$~~~~~\hat{\hat{P}}~~~~~$ \scriptsize ($\tdc{}s$, choices and holes)};
\node [block, right of=p1, text width=7.2em, node distance=4cm] (p2) {Type-Directed Expansion Rules};
\node [text centered, right of=p2, text width=3.8em, node distance=3.5cm] (p3) {$~~~~\hat{P}~~~~$ \scriptsize (choices and holes)};
\node [block, right of=p3, text width=5.2em, node distance=3cm] (p4) {Constraint-based Synthesis};
\node [text centered, right of=p4,text width=5.2em, node distance=3cm] (p5) { $~~~~\overline{P}~~~~$ \scriptsize (no synthesis constructs)};

\path [line] (p1) edge  (p2);
\path [line] (p2) edge  (p3);
\path [line] (p3) edge  (p4);
\path [line] (p4) edge  (p5);

\end{tikzpicture}
\end{adjustbox}
\vspace{0.1cm}

First, \lang{} uses a set of expansion rules that uses bi-directional type checking to eliminate the \tdc{}s.  The result  is a program that only contains choices and holes. The second step is to use a constraint-based approach to solve for these choices. The next subsections will present each of these steps in more detail.


\subsection{Type-Directed Expansion Rules}
We will now formalize the process of specializing  and expanding the \tdc{}s into sets of possible expressions. 
We should first note that the expansion and the specialization of the different \tdc{}s interact in complex ways.
 For example, for the \casehc{} construct in the running example, the system cannot determine which cases to generate until it knows the type of \C{src}, which is only fixed once the \emph{polymorphic generator} for \C{recursiveReplacer}  is specialized to the calling context. On the other hand, if a \emph{polymorphic generator} is invoked inside the body of a \casehc{} (like \C{rcons} in the running example), we may not know the types of the arguments until after the \casehc{} is expanded into separate cases. Because of this, type inference and  expansion of the \tdc{}s must happen in tandem. 


We formalize the process of expanding \tdc{}s using two different kinds of judgements. The \emph{typing judgement}  $\Gamma\vdash e:\theta$ determines the type of an expression by propagating information bottom-up from sub-expressions to larger expressions.  
On the other hand, \tdc{}s cannot be type-checked in a bottom-up manner; instead, their types must be inferred from the context.
The \emph{expansion judgment}  $\Gamma\vdash e~\xrightarrow{\theta }~e'$ expands an expression  $e$ involving \tdc{}s into an expression $e'$ that does not contain \tdc{s} (but can contain choices and holes).
In this judgment, $\theta$ is used to propagate information top-down and represents the type required in a given context; 
in other words, after this expansion, the typing judgement $\Gamma\vdash e' : \theta$ must hold. 
We are not the first to note that bi-directional typing~\cite{PierceBidirectional} can be very useful in pruning the search space for synthesis~\cite{OseraZ15,synquid}, but we are the first to apply this in the context of constraint-based synthesis and in a language with user-provided definitions of program spaces. 

\begin{figure}
\centering
\begin{adjustbox}{max width=\textwidth}
$\begin{array}{c}
FUN~~~
\frac{\begin{array}{c}\Gamma;\left \{x_{i}:\tau _{i}\right \}_{i<  n}\vdash e~\xrightarrow{\tau _{o}}~e'\end{array}}
{\begin{array}{c}\Gamma\vdash \tau _{o}~f \left(\left \{x_{i}:\tau _{i}\right\}_{i<  n}\right)e
\xrightarrow{\perp}\tau _{o}~f \left(\left \{x_{i}:\tau _{i}\right \}_{i<  n}\right)~e'\end{array}} 
\\
~\\
FL~~~
\frac{~~\begin{array}{c}
\Gamma\vdash e:\left \{l_{i}:\tau _{i}\right \}_{i < n}~~~
\Gamma\vdash e~\xrightarrow{\{l_{i}:\tau _{i}\}_{i<n}}e^{'}~~~~
\{\tau_{i_j} = \tau_0\}_j~~~~~
~~(\tau _{0}[~]=\tau)~~~
\end{array}}{\begin{array}{c}\Gamma\vdash e.\mc{fields?}{} \xrightarrow{\tau }~\{\{e^{'}.l_{i_{j}} \}_j\}\end{array}} 
\\
~\\
FPM~~
\frac{~~\begin{array}{c}
\Gamma=\left (\Gamma^{'};x~:\sum~name_{i}\left \{l_{k}^{i}:\tau _{k}^{i}\right \}_{k<n_i}\right )~~~~~~~~~~~~
\left\{\left (\Gamma^{'};x~:\left \{l_{k}^{i}:\tau _{k}^{i}\right \}_{k<n_i}\right)\vdash ~e~\xrightarrow{\theta }~~e_{i}\right\}_i \\
\end{array}~}{\begin{array}{c}
\Gamma\vdash \mc{switch}\left(x\right)\left \{~\mc{case?}{}:e~\right \} 
\xrightarrow{\theta } 
\mc{switch}\left(x\right)\left \{~\mc{case}~name_{i}:e_{i}\right \}_i\\
\end{array}~~}\\
~\\
UC1~~
\frac{\begin{array}{c}
\tau =\Sigma name_{i}\left \{l_{k}^{i}:\tau_{k}^{i}\right \}_{k < n_i} ~~~~~~~~~~~~~~
 e_{1}~\xrightarrow{\tau_{k}^{i}}~~e_{1_k}^{i} \ldots 
 e_{m}~\xrightarrow{\tau_{k}^{i}}~~e_{m_k}^{i}\\
\end{array}}{\begin{array}{c}
\Gamma\vdash ~\mc{new}~\mc{cons?}{}\left(e_1\ldots e_m\right)~
\xrightarrow{\tau}~
\mc{choose}\left(\left \{\mc{new}~name_{i}\left(\left \{l_{k}^{i}=\mc{choose}\left(\{e_{r_k}^{i}\}_{r<m}\right)\right \}_{k < n_i} \right)\right \}_i\right)
\end{array}} \\
~\\
UC2~~
\frac{\begin{array}{c}\tau = prim \end{array}}
{\begin{array}{c}\Gamma\vdash ~~\mc{new}~\mc{cons?}{}\left(e_1\ldots e_m\right)~
\xrightarrow{\tau}~
??
\end{array}} \\
~\\
PG~~~
\frac{\begin{array}{c}
\theta _{out}~~\hat{f}\left\langle \left \{T_{i}\right \}\right\rangle \left(\left \{p_{i}:\theta _{i}\right \}_i\right)~e~~~~~~~~~~~~~~~~
\Gamma\vdash e_{i}:\tau_{i}^{in}~~~{for~i<  k}\\
S=\mc{Unify}\left (\left \{(\theta _{out},~\theta)\right \}\cup \left \{(\theta _{i},\tau_{i}^{in})\right \}_{i<  k}\right ) \\
\begin{array}{c}
e_{i}\xrightarrow{S(\theta _{i})}e_{i}'~~~{for~ i\leq k+n} ~~~~~~~~~~~~~~~~
e[\{e_i'/p_i\}_i] \xrightarrow{S(\theta)}~e'\\
\end{array}~~~ \\
\end{array}}{\begin{array}{c}\hat{f}\left(e_{0}\ldots  e_{k}\ldots  e_{k+n}\right)\xrightarrow{\theta }~e' \end{array}} \\
\end{array}$
\end{adjustbox}
\caption{Expansion rules for various language constructs}
\vspace{-15pt}
\figlabel{exprules}
\end{figure}

The expansion rules for functions and \tdc{}s are shown in \figref{exprules}. At the top level, given a program  $P$,  every function in  $P$  is transformed using the expansion rule FUN.
The body of the function is expanded under the known output type of the function. The most interesting cases in the definition of the expansion judgment correspond to the \tdc{}s as outlined below. The  expansion rules for the other expressions are straightforward and are elided for brevity. 

\noindent
\textbf{\textit{Field List}}
The rule FL shows how a  \emph{field list} is expanded. If the required type 
is an array of $\tau_0$, then this \tdc{} can be expanded into an array of all fields of type $\tau_0$.

\noindent
\textbf{\textit{Flexible Pattern Matching}}
For each case, the body of \casehc{} is expanded while setting $x$ to  a different type corresponding to each variant $name_{i}\left \{l_{k}^{i}:\tau_{k}^{i}\right \}_{k<n_i}$ as shown in the rule FPM. Here, the argument to \C{switch} is required to be a variable so that it can be used with a different type inside each of the different cases. 
Note that each case is expanded independently, so the synthesizer can make different choices for each $e_{i}$.

\noindent
\textbf{\textit{Unknown constructor}}
If the required type 
is an ADT, 
the rule UC1 expands the expressions passed to the \emph{unknown constructor} based on the type of each field of each variant of  the ADT
and uses the resulting 
expressions to initialize the fields in the relevant constructor. It returns a \C{choose} expression with all these constructors as the arguments. If the required type is a primitive type (int or bit), 
the unknown constructor is expanded into a \Sk{} hole by the rule UC2. 

\noindent
\textbf{\textit{Polymorphic Generator Calls}}
When the expansion encounters a  call to a \emph{polymorphic generator}, the generator will be expanded and specialized according to the PG rule. 
When a generator is called with arguments  $\{e_{i}\}_i$,  we can  separate the arguments into expressions that can be typed using the standard typing judgement, and expressions such as \C{new}~\conshc{}\C{(...)} that cannot. In the rule, we assume, without loss of generality, that the first  $k$  expressions can be typed and the reminder cannot.
The basic idea behind the expansion is as follows. First, the rule obtains the types of the first  $k$ arguments and unifies them with the types of the formal parameters of the function to get a  type substitution  $S$.
The arguments to the original call are expanded with our improved knowledge of the types, and the body of the generator is then inlined and expanded in turn. 
The actual implementation also keeps track of how many times each generator has been inlined and replaces the generator invocation with \C{assert false} when the inlining bound has been reached.

The above expansion rules fail if a type variable is encountered in places where a concrete type is expected, and in such cases the system will throw an error. For example, expressions such as \C{field(field(e))}, where \C{field} is as defined in \figref{tmp-sol}, cannot by type-checked in our system because the expected type of the inner \C{field} call cannot be determined using top-down type propagation.  

\subsection{Constraint-based Synthesis}\seclabel{synthesis}

Once we have a program with a fixed number of integer unknowns, the synthesis problem can be encoded as a constraint 
$
\exists \phi .~\forall \sigma .~P(\phi, \sigma) 
$ where  $\phi $  is a \emph{control vector} describing the set of choices that the synthesizer has to make,  $\sigma $  is the input state of the program, and  $P(\phi, \sigma)$  is a predicate that is true if the program satisfies its specification under input  $\sigma $  and control $\phi$.
Our system follows the standard approach of unrolling loops and inlining recursive calls to derive $P$ and uses counterexample guided inductive synthesis (CEGIS) to solve this doubly quantified problem~\cite{sketchthesis}. For readers unfamiliar with this approach, the most relevant aspect from the point of view of this paper is that the doubly quantified problem is reduced to a sequence of inductive synthesis steps. At each step, the system generates a solution that works for a small set of inputs, and then checks if this solution is in fact correct for all inputs; otherwise, it generates a counter-example for the next inductive synthesis step.

Applying the standard approach can, however, be problematic in our context especially with regards to inlining recursive calls. For instance, consider the example from \secref{overview}. Here, the function \C{desugar} that has to be synthesized is a recursive function. If we were to inline all the recursive calls to \C{desugar}, then a given concrete value for the input $\sigma$ such as \C{BetweenS(a = NumS(...), b = BinaryS(...), ...)},  will exercise multiple cases within \C{desugar} (\C{BetweenS}, \C{NumS} and \C{BinaryS} for the example). This is problematic in the context of CEGIS, because at each inductive synthesis step the synthesizer has to jointly solve for all these variants of \C{desugar} which greatly hinders scalability  when the source language has many variants. 

\subsection{Inductive Decomposition}\seclabel{optim}
The goal of this section is to leverage the inductive specification to potentially avoid inlining the recursive calls to the synthesized function. 
This idea of treating the specification as an inductive hypothesis is well known in the deductive verification community where the goal is to solve the following problem: $\forall \sigma. ~P(\phi_0, \sigma)$. However, in our case, we want to apply this idea during the inductive synthesis step of CEGIS where the goal is to solve $\exists \phi.~P(\phi, \sigma_0)$ which  has not been explored before.

\begin{definition}[Inductive Decomposition]
		Suppose the specification is of the form $interp_s(e)  = interp_d(trans(e))$ where $trans$ is the function that needs to be synthesized. Let $trans(e')$ be a recursive call within $trans(e)$ where $e'$ is strictly smaller term than $e$. Inductive Decomposition is defined as the following substitution: 
		1. Replace $trans(e')$ with a special expression $\boxed{e'}$. 
		2. When inlining function calls, apply the following rules  for the evaluation of $\boxed{e'}$:
		
		\centering
			$\begin{array}{c}
				interp_d(\boxed{e'}) \xrightarrow{~~~~~~~~} interp_s(e') \\
				\boxed{e'} \text{ in any other context } \xrightarrow{~~~~~~~~} trans(e')
			\end{array}$
	
\end{definition}
i.e.  Inductive Decomposition works by delaying the evaluation of a recursive $trans(e')$ call by replacing it with a placeholder that tracks the  input $e'$. Then, if the algorithm encounters these placeholders when inlining $interp_d$ in the specification,  
it replaces them directly with $interp_s(e')$  which we know how to evaluate, thus, eliminating the need to inline the unknown $trans$ function. This replacement is sound because the specification states $interp_d(trans(e')) = interp_s(e')$.
If the algorithm encounters the placeholders in any other context where the inductive specification can not be leveraged, it defaults to evaluating $trans(e')$.


\begin{theorem}
	Inductive Decomposition is sound and complete.
	In other words, if the specification is valid before the substitution, then it will be valid after the substitution and vice-versa.
\end{theorem}

A proof of this theorem can be found in the tech report~\cite{techreport}. Although the Inductive Decomposition algorithm imposes restrictions on which recursive calls can be eliminated, it turns out that for many of the ADT transformation scenarios, the algorithm can totally eliminate all recursive calls to $trans$. For instance, in the running example, because of the inductive structure of \C{dstInterpret}, all placeholders for recursive \C{desugar} calls  will occur only in the context of \C{dstInterpret(desugar(e'))} which can be replaced by \C{srcInterpret(e')} according to the algorithm. Thus, after the substitution, the \C{desugar} function is no longer recursive and moreover, the desugaring for the different variants can be synthesized separately. For the running example, we gain a 20X speedup using this optimization. 
Our system also implements several generalizations of the aforementioned optimization
that are detailed in the tech report~\cite{techreport}.

\newcommand{\rot}[2]{\parbox[t]{2mm}{\multirow{#2}{*}{\rotatebox[origin=c]{90}{#1}}}}
\newcommand*\rott{\rotatebox{90}}

\section{Evaluation}\seclabel{experiments}
\paragraph{\textbf{Benchmarks}}
We evaluated our approach on 23 benchmarks as shown in \figref{benchmarks}. All  benchmarks along with the synthesized solutions can be found in the tech report~\cite{techreport}. Since there is no standard benchmark suite for morphism problems, we chose our benchmarks from common assignment problems (the lambda calculus ones), desugaring passes from  \Sk{} compiler and  some standard data structure manipulations on trees and lists. The AST optimization benchmarks are from  a system that synthesizes simplification rules for SMT solvers~\cite{rohit}. 

\paragraph{\textbf{Templates}} 
The templates for all our benchmarks  use  one of the four generic descriptions  we have in the library. All benchmarks except \C{arrAssertions}, \C{NegNorm} and  AST optimizations  use a generalized version of the \C{recursiveReplacer} generator seen in \figref{tmp-sol} (the exact generator is in the tech report).  This generator is also used as a template for problems that are very  different from the desugaring benchmarks such as the list and the tree manipulation problems, illustrating how generic and reusable the templates can be.  
The \C{arrAssertions} benchmark differs slightly from the others as its ADT definitions  have arrays of recursive fields and hence, we have a version of the recursive replacer that also recursively iterates over these arrays. The \C{NegNorm} benchmark requires a template that has nested pattern matching.
Another interesting example of reusability of templates is the AST optimization benchmarks. All 5 benchmarks in this category are synthesized from a single library function. 
The \C{template} column in \figref{benchmarks} shows the number of lines used in the template for each benchmark. Most benchmarks have a single line that calls the appropriate library description  similar to the  example in \secref{overview}. Some benchmarks  also specify additional components such as helper functions that are required for the transformation. Note that these additional components will also be required for other systems such as Leon and Synquid.

\subsection{Experiments}
\paragraph{\textbf{Methodology}}
All  experiments were run on a machine with forty 2.4 GHz Intel Xeon processors and 96GB RAM. We ran each experiment 10 times and report the median. 

\noindent
\textbf{\textit{Hypothesis 1: Synthesis of complex routines is possible}}
\figref{benchmarks} shows the running times for all our benchmarks  (\C{T-opt} column). \lang{} can synthesize all but one benchmark very efficiently when run on a single core using less than 1GB memory---19 out of 23 benchmarks take $\leq$ 1 minute. 
  Many of these benchmarks are beyond what can be synthesized by other tools like Leon, Rosette, and others and yet, \lang{} can synthesize them just from very general templates. For instance, the \C{lcB} and \C{lcP} benchmarks are automatically discovering the Church encodings for boolean operations and pairs, respectively. The  \C{tc} benchmark synthesizes an algorithm to produce type constraints for lambda calculus ASTs to be used to do type inference. The output of this algorithm is a conjunction of type equality constraints which is produced by traversing the AST. Several  other desugaring benchmarks have specifications that involve complicated interpreters that keep track of state, for example. Some of these specifications are even undecidable and yet, \lang{} can synthesize these benchmarks (up to bounded correctness guarantees). The figure also shows the size of the synthesized solution (\C{code} column)\footnote{Solution size is measured as the number of nodes in the AST representation of the solution}. 
  
There is one benchmark (\C{langState}) that cannot be solved by \lang{} using a single core. Even in this case, \lang{} can synthesize the desugaring for 6 out of 7 variants in less than a minute. The unresolved variant requires generating expression terms that are very deep which exponentially increases the search space.
Luckily, our solver is able to leverage multiple cores using the random concretization technique~\cite{adaptCon} to search the space of possible programs in parallel. 
The column \C{T-parallel} in \figref{benchmarks} shows  the running times for all benchmarks when run on 16 cores. \lang{} can now  synthesize all variants of the \C{langState} benchmark in about 9 minutes. 

The results discussed so far are obtained for optimal search parameters for each of the benchmarks.
We also run an experiment to randomly search for these parameters using the parallel search technique with 16 cores and report the results in the \C{T-search} column. 
Although these times are higher than when using the optimal parameters for each benchmark (\C{T-parallel} column), the difference is not huge for most benchmarks. 

\begin{figure}[t]
\footnotesize
\centering
\begin{adjustbox}{max width=\textwidth}
\begin{tabular}{ |c| l | l | c c | c  c  c  c |}
\hline 
 & Bench & Description & template & code & T-opt & T-parallel  & T-search  & T-unopt \\
  \hline
  \rot{Desugar}{9} 
  & {lang} & {Running example}& 1 & 50 &  {7.5}   & {8.6}   &  {85.9} & {152.5} \\
  & {langState} & {Running example with mutable state} & 1 &62&  {$\bot$} & {527.2} &  {1746.9} & {$\bot$} \\
  & {regex} & {Desugaring regular expressions} & 1 & 22 & {2.0} & {3.3}  &  {9.1} & {3.3}\\
  & {elimBool} & {Boolean operations to if else}  & 1 & 21&  {1.5} & {2.9} & {7.5} & {2.4}  \\
  & {compAssign} & {Eliminates compound assignments} & 1 & 42& {16.6} & {20.9}  & {31.8} & {176.2}  \\
  & {langLarge} & {Desugaring a large language} & 1 & 126& {61.2} & {58.0} & {49.7} & {$\bot$} \\
  & {arrAssertions} & {Add out of bounds assertions}  & 3 & 40& {37.2} & {50.5}  & {66.7}  &{53.0} \\
  & {NegNorm} & {Computes negation normal form} & 3 & 57 & 21.2 & 13.6 & 64.4 & $\bot$ \\
  & {lcB} & {Boolean operations to $\lambda$-calculus}  & 1 & 55& {43.1} & {47.4}  & {40.6} & {43.1} \\
  & {lcP} & {Pairs to $\lambda$-calculus} & 1 & 41& {163.6} & {258.2} & {288.3}  & {163.6}  \\
  \hline
  \rot{Analysis}{3} 
  &&&&&&&&\\
  & {tc} & {Type constraints for $\lambda$-calculus} & 8 & 41& {168.9} & {68.0} & {201.9} & {168.9}  \\
    &&&&&&&&\\
  \hline
\rot{AST optim}{5} 
  & {andLt} & {AST optimization 1} & 1 & 15 & {3.1} & {3.1} & {13.2} & {N/A}\\
  & {andNot} & {AST optimization 2}  & 1 & 6& {2.6} & {3.0}  & {13.0} & {N/A} \\
  & {andOr} & {AST optimization 3} & 1 & 12 & {3.7} & {3.1}  & {14.0}  & {N/A}\\
  & {plusEq} & {AST optimization 4}  & 1 & 18&{3.3} & {3.0} & {14.0} & {N/A} \\
  & {mux} & {AST optimization 5}  & 1 & 6& {2.4} & {3.0} & {12.4} & {N/A}\\
  \hline
  \rot{List}{3}
  & {lIns} & {List insertion} & 1 & 12& {1.5} & {2.3} & {2.2} & {2.1} \\
  & {lDel} & {List deletion} & 2 & 14 & {4.0} & {4.6} & {4.1} & {3.1} \\
  & {lUnion} & {Union of two lists}  & 1 & 10& {8.7} & {2.7}& {4.8} & {2.1} \\
  \hline
  \rot{Tree}{4}
  & {tIns} & {Binary search tree insertion}  & 1 & 48 & {20.7} & {14.5}  & {41.6} & {11.6}  \\
  & {tDel} & {Binary search tree deletion}& 4 & 63 & {224.8} & {227.4}  & {286.1}  & {298.9} \\
   & {tDelMin} & {Binary search tree delete min}  & 2 & 18& {27.1} & {32.2} & {57.7} & {24.9}\\
    & {tDelMax} & {Binary search tree delete max} & 2 & 18& {25.9} & {30.8} & {54.4} & {25.9}\\
 \hline
\end{tabular}
\end{adjustbox}
\caption{Benchmarks. All reported times are in seconds. $\bot$ stands for timeout (> 45 min) and N/A stands for not applicable.}
\vspace{-15pt}
\figlabel{benchmarks} 
\end{figure}

\noindent
\textbf{\textit{Hypothesis 2: The Inductive Decomposition  improves the scalability.}}
In this experiment, we run each benchmark with the \emph{Inductive Decomposition} optimization disabled  and the results are shown in \figref{benchmarks} (\C{T-unopt} column). This experiment is run on a single core. 
First of all, the technique is not applicable for the AST optimization benchmarks because the functions to be synthesized are not recursive.  Second, for three benchmarks---the $\lambda$-calculus ones and the \C{tc} benchmark, we noticed that their specifications do not have the inductive structure and hence, the optimization never gets triggered. 

But for the other benchmarks, it can be seen that \optname{} leads to a substantial speed-up on the bigger benchmarks. Three benchmarks time out (> 45 minutes) and we found that \C{langState} times out even when run in parallel. In addition, without the optimization, all the different variants need to be synthesized together and hence, it is not possible to get partial solutions. The other benchmarks show an average speedup of 2X  with two benchmarks having a speedup > 10X. We found that for benchmarks that have very few variants, such as the list and the tree benchmarks, both versions perform almost similarly.

To evaluate how the performance depends on the number of variants in the initial AST, we considered the \C{langLarge} benchmark that synthesizes a desugaring for a source language with 15 variants into a destination language with just 4 variants. We started the benchmark with  3 variants in the source language while incrementally adding the additional variants and measured the run times both with the optimization enabled and disabled. The graph of run time  against the number of variants is shown in \figref{optimGraph}. It can be seen that without the optimization the performance degrades very quickly and moreover, the unoptimized version times out (> 45 min) when the number of variants is > 11.

\subsection{Comparison to other tools}\seclabel{comp}
We compared \lang{} against three tools---Leon, Synquid and Rosette that can express our benchmarks. The list and the tree benchmarks are the typical benchmarks that Leon and Synquid can solve and they are faster than us on these benchmarks.  However, this difference is mostly due to \lang{}'s final verification time. For these benchmarks, our verification is not at the state of the art because we use a very naive library for the \C{set} related functions used in their specifications.  We also found that Leon and Synquid can synthesize some of our easy desugaring benchmarks that requires constructing relatively small ADTs like \C{elimBool} and \C{regex} in almost the same time as us. However, Leon and Synquid were not able to solve the harder desugaring problems including the running example. We should also note that this comparison is not totally apples-to-apples  as Leon and Synquid are more automated than \lang{}. 

\begin{wrapfigure}{r}{0.5\textwidth}
	\vspace{-0.2cm}
	\centering
	\includegraphics[width=.5\textwidth]{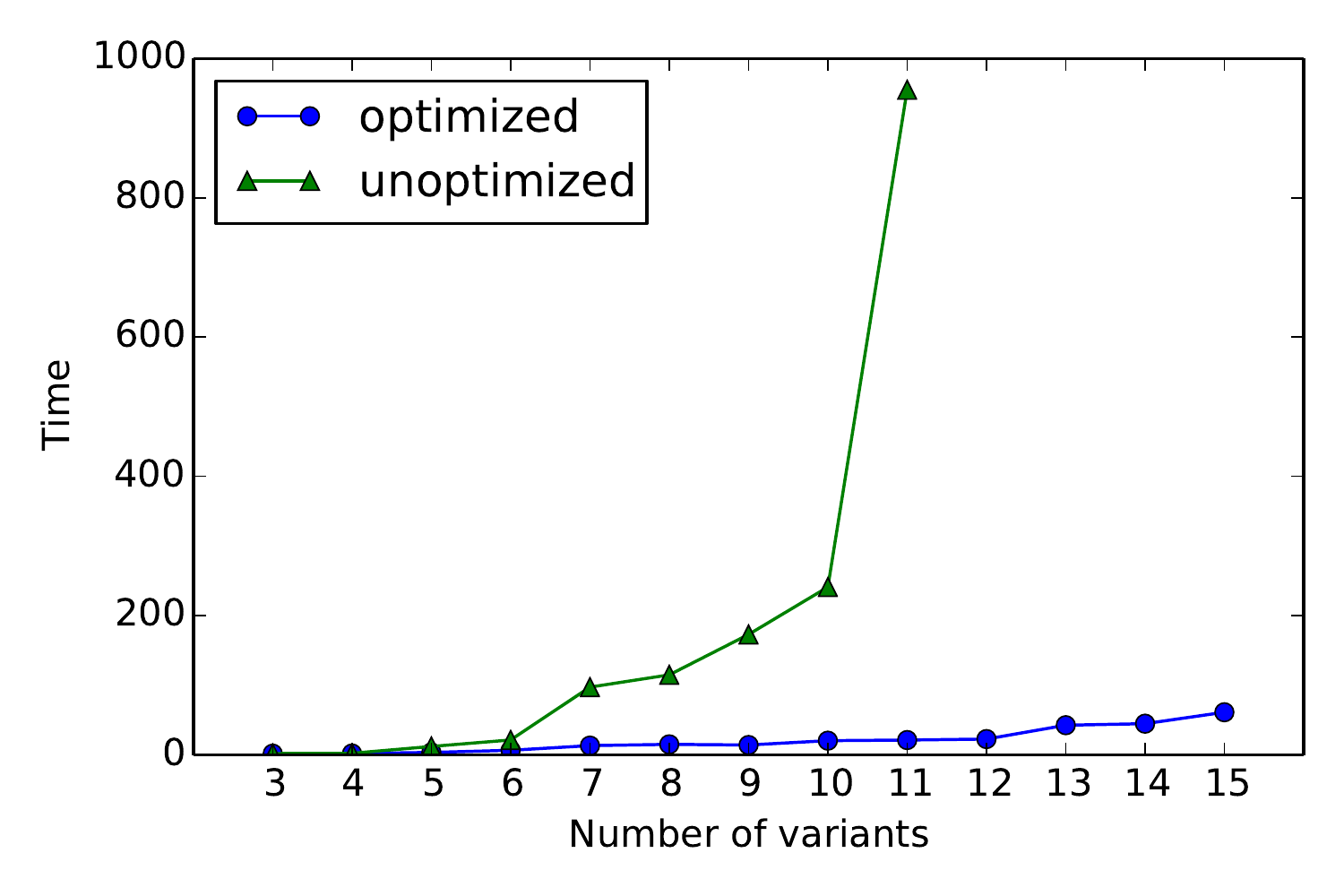}
	\caption{Run time (in seconds) versus the number of variants of the source language for the \C{langLarge} benchmark with and without the optimization.}
	\figlabel{optimGraph}
	\vspace{-0.2cm}
\end{wrapfigure}

For comparison against Rosette, we should first note that since Rosette is also a SyGus solver, we had to write very verbose templates for each benchmark. 
But even then, we found that Rosette cannot get past the compilation stage because the solver gets bogged down by the large number of recursive calls requiring expansion. 
For the other smaller benchmarks that were able to get to the synthesis stage, we found that Rosette is either comparable or slower than \lang{}. For example, the benchmark \C{elimBool} takes about 2 minutes in Rosette compared to 2s in \lang{}. We attribute these differences to the different solver level  choices made by Rosette and \Sk{} (which we used to built \lang{} upon). 


\section{Related Work}\seclabel{relwork}
There are many recent systems that synthesize recursive functions on algebraic data-types.
  Leon~\cite{Leon,KneussKKS13,Kuncak14} and Synquid~\cite{synquid} are two systems that are very close to ours. 
Leon, developed by the LARA group at EPFL, is built on prior work on  complete functional synthesis  by the same group~\cite{completefunctionalsynthesis}
and moreover, their recent work on Synthesis Modulo Recursive Functions~\cite{KneussKKS13} demonstrated a sound technique to synthesize provably correct recursive functions involving algebraic data types.
Unlike our system, which relies on bounded checking to establish the correctness of candidates, their procedure is capable of synthesizing provably correct implementations. The tradeoff is the scalability of the system; Leon supports using arbitrary recursive predicates in the specification, but in practice it is limited by what is feasible to prove automatically. Verifying something like equivalence of lambda interpreters fully automatically is prohibitively expensive, which puts some of our benchmarks beyond the scope of their system. 
 Synquid~\cite{synquid}, on the other hand,  uses refinement types as a form of specification to efficiently synthesize programs. Like our system, Synquid also depends on bi-directional type checking to effectively prune the search space. But like Leon, it is also limited to decidable specifications.
There has also been a lot of recent work on programming by example systems for synthesizing recursive programs~\cite{OseraZ15,FeserCD15,AlbarghouthiGK13,PerelmanGGP14}.  All of these systems rely on  explicit search  with some systems like~\cite{OseraZ15} using bi-directional typing to prune the search space and other systems like~\cite{AlbarghouthiGK13}  using specialized data-structures to efficiently represent the space of implementations. 
However, they are limited to programming-by-example settings, and cannot handle our benchmarks, especially the desugaring ones.

Our work builds on a lot of previous work on SAT/SMT based synthesis from templates. Our implementation itself is built on top of the open source Sketch synthesis system~\cite{sketchthesis}. However, several other solver-based synthesizers have been reported in the literature, such as Brahma~\cite{sumitbitvector}. 
More recently, the work on the solver aided language Rosette~\cite{TorlakB14,TorlakB13} has shown how to embed synthesis capabilities in a rich dynamic language and then how to leverage these features to produce synthesis-enabled embedded DSLs in the language. Rosette is a very expressive language and in principle can express all the benchmarks in our paper. However, Rosette is a dynamic language and lacks static type information, so in order to get the benefits of the high-level synthesis constructs presented in this paper, it would be necessary to re-implement all the machinery in this paper as an embedded DSL. 

There is also some related work in the context of using polymorphism to enable re-usability in programming.  \cite{scrap} is one such approach  where the authors describe a design pattern in Haskell that allows programmers to express  the boilerplate code required for traversing recursive data structures in a reusable manner. This paper, on the other hand, focuses on supporting reusable templates in the context of synthesis which has not been explored before. 
Finally, the work on hole driven development~\cite{agda} is also related in the way it uses types to gain information about the structure of the missing code. The key difference is that existing systems like Agda lack the kind of symbolic search capabilities present in our system, which allow it to search among the exponentially large set of expressions with the right structure for one that satisfies a deep semantic property like equivalence with respect to an interpreter.

\section{Conclusion}\seclabel{conclusion}
The paper has shown that by combining type information from algebraic data-types together with the novel Inductive Decomposition optimization, it is possible to efficiently synthesize complex functions based on pattern matching from very general templates, including desugaring functions for lambda calculus that implement non-trivial Church encodings. 

\noindent
\textbf{Acknowledgments:} We would like to thank the authors of Leon and Rosette for their help in comparing against their systems and the reviewers for their feedback. This research was supported by NSF award \#1139056 (ExCAPE).




\bibliographystyle{abbrv}
\bibliography{bibliography}
\newpage
\appendix
\section{Benchmarks and Library Components}
All the benchmarks along with corresponding synthesized solutions can be found at https://bitbucket.org/jeevana\_priya/syntrec-benchmarks/src. The library generators and components are  in the lib.skh file in the above repository.
The solutions generated by \lang{} are a little verbose because of the temporary variables and also beacuse the function outputs are converted into reference parameters. 

\section{Static Semantics of \lang{} language}

The typing rules for the language work as one would expect. For example, the type of a field access from a record is the type of the field. 
\begin{formula}
\frac{\Gamma\vdash e:\{\{l_{i}:\tau_{i}\}_i \}~~~~~~l=l_{i}~~~~~\tau =\tau_{i}}{\Gamma\vdash e.l:\tau }
\end{formula}
Values in the ADT are created through constructors. 
\begin{formula}
\frac{\begin{array}{c}
\tau_{adt}=\Sigma name_{i}\left \{\{l_{k}^{i}:\tau_{k}^{i}\}_{k < n_i}\right \} 
~~~name=name_{t}~~~~~\{l_{k} = l_{k}^{t}~~~~~~~\Gamma\vdash e_{k}:\tau_{k}^{t}\}_{k < n_t} \\
\end{array}}{\Gamma\vdash new~name(l_{j}=e_{j}):\tau_{adt}}
\end{formula}

The most interesting rule is the one for $switch$. The rule, shown below, assumes that the argument  $x$  to  $switch$  is a variable whose type is an ADT where each variant corresponds to one of the cases in the switch. The body of each case is then type checked under the assumption that the type of  $x$  is the type associated with the corresponding variant.
\begin{formula}
~\frac{\begin{array}{c}
\Gamma=(\Gamma^{'};x:\tau_{adt})~~~~~~\tau_{adt}=\Sigma name_{i}\left \{\{l_{k}^{i}:\tau_{k}^{i}\}_{k < n_i}\right \} \\
~\{(\Gamma^{'};x:\left \{\{l_{k}^{i}:\tau_{k}^{i}\}_{k < n_i}\right \})\vdash ~e_{i}:\tau\}_i  \\
\end{array}}
{\Gamma~\vdash ~switch\left(x\right)\left \{~case~name_{i}:e_{i}~~\right \}_i:\tau }~
\end{formula}

\section{Dynamic Semantics of \lang{} language}
The dynamic semantics evaluate expressions under an environment  $\sigma $  that tracks the values of variables. ADT values are represented with a named record  $\left\langle name,~\{l_{i}=v_{i}\}_i\right\rangle $  that has the name of the corresponding variant and the values for each field. This is illustrated by the rule for the constructor.
\begin{eqnarray*}
\frac{\{\sigma ,e_{i}\rightarrow v_{i}\}_i~~~~~v=\left\langle name,~\{l_{i}=v_{i}\}_i\right\rangle }{\sigma ,new~name\left(\{l_{i}=e_{i}\}_i\right)\rightarrow v} 
\end{eqnarray*}
The name stored as part of the record is used by the switch statement in order to choose which branch to evaluate.
\begin{eqnarray*}
\frac{\sigma \left(x\right)=\left\langle name,~\{l_{i}=v_{i}\}_i\right\rangle ~~~name_{j}=name~~~~~~\sigma ,e_{j}\rightarrow v}{\sigma ,switch\left(x\right)\left \{~case~name_{i}:e_{i}\right \}_i\rightarrow v} 
\end{eqnarray*}
and it is  easy to show that the typing rules ensure that the field access rule below will always find a matching field  $l_{i}=l$ .
\begin{eqnarray*}
\frac{\sigma ,e\rightarrow \left\langle name,~\{l_{i}=v_{i}\}_i\right\rangle ~~~l=l_{i}~~~v=v_{i}}{\sigma ,~e.l\rightarrow v} 
\end{eqnarray*}

\section{Proof of Inductive Decomposition theorem}

The proof has two parts; first we show that the substitution is complete, meaning that if the original specification is valid for a given synthesized  $trans$  function, then it will still be valid after we perform the substitution. Second, we show that the substitution is sound i.e. if the original specification does not hold for a given $trans$, then it will also not hold after the substitution. Also note that we can assume that there is only one recursive call ($trans(e')$) that undergoes the transformation, because if replacing one call has no effect, then the calls can be replaced one by one until all the calls have been replaced. 

\noindent
\textbf{Completeness:}
We start by assuming that the specification is indeed valid for a particular $trans$. In this case, $interp_d(\boxed{e'})$ is in fact equivalent to $interp_s(e')$ (since the specification is valid for all $e$). Hence, the specification is still valid after the substitution.

\noindent
\textbf{Soundness:}
Let us assume that the specification does not hold, and let  $e_{x}$  be the smallest ADT value such that  $interp_{s}\left(e_{x}\right)\neq  interp_{d}(trans\left(e_{x}\right))~$. We define smallest in terms of the maximum depth of recursion of  $e_{x}$  (the height of the tree, if we think of  $e_{x}$  as a tree). Now, if $e_{x}$ is not recursive,  we are done since the substitution will only affect recursive values. Now, if $e_x$ is recursive and assume that there is a recursive $trans(e_x')$ that gets replaced by the substitution. Here, since $e_x'$ is smaller than $e_x$ and because $e_x$ is the smallest ADT that does not satisfy the specification, $interp_d(\boxed{e_x'})$ is still equivalent to $interp_s(e_x')$.  Hence, the substitution has no impact on validity of $e_x$ and so, $e_x$ will also fail the specification.

Note that the proof of soundness above only works because the recursive calls inside $trans$ operate on trees that are smaller than the input tree. This is an important condition that
is usually enforced by the template; if it were not to hold, the transformation could take a buggy implementation (one that has an infinite recursion, for example) and make it appear correct. 

\section{Structural constraints for full application of Inductive Decomposition}
Below, we  first define the several structural constraints such that if these constraints were to hold, then the inductive decomposition will \emph{totally eliminate} all recursive $trans$ calls. 

\begin{definition}[Recursive Transformer]
\\
Given an ADT $\tau =\sum_{i}Q_{i}\left \{l_{j}^i:\tau_{j}^i\right \}_{j<n_i}$, a function $f$ is a \textbf{Recursive Transformer} if it has the following form: 
\begin{formulaarr}
f\left(e\right):= 
switch\left(e\right)\{case~Q_i:proc^{Q_i}(
 \{f(e.l_{j_k}^i)\}_{k<b_i}, 
 \{e.l_{j'_k}^i \}_{k<b'_i}) \}
\end{formulaarr}
Where none of the fields $e.l_{j'_k}^i$ are of type $\tau$ (but all the fields $e.l_{j_k}^i$ must be of type $\tau$ for the recursive call to be well typed).
\end{definition}

In other words, $f$ will pattern match on $e$, and in each case it will make recursive calls on certain fields of  $e$  and process the results through an arbitrary function  $proc^{Q_i}$  before returning them. 

\begin{definition}[Recursive Morphism]
\\
A \textbf{Recursive Morphism} is a \textbf{Recursive Transformer} with the following additional constraint:

$proc^{Q_i}(v_0,\ldots v_k, v'_0, \ldots, v'_k)=\psi[v_0,\ldots v_k]$ where $v_0 \ldots v_k$ are the terms involving the recursive calls to this  morphism function and $\psi$ is an expression that constructs an ADT tree (potentially of a different type than $\tau$) with $v_0 \ldots v_k$ as some of the leaves 
and $\psi$ itself does not depend on them, so \\$proc^{Q_i}(u_0,\ldots u_k,~ v'_0, \ldots, v'_k)=\psi[u_0,\ldots u_k]$ with the same $\psi$.
\end{definition}

For example, the \C{desugar} function is a recursive morphism where $\psi$  for each case is the unknown expression tree of type \C{dstAST} that is generated by the \C{rcons} function in the template (line 6 in \figref{tmp-sol}).   
On the other hand, the interpreters in the running example that compute the value of an expression from the values of its sub-expressions are examples of  recursive transformers by the above definition, but are not  morphisms because they read the recursively evaluated values.

\begin{lemma}
Let $interp_s$, $interp_d$ be two \textit{recursive transformers} that operate on two recursive data types $\tau _{s}=\sum_{i}K_{i}\left \{l_{j}^i:\tau^{s^i} _{j}\right \}_{j<n_i}$ and 
$\tau _{d}=\sum_{i}Q_{i}\left \{l_{j}^i:\tau _{j}^{d^i}\right \}_{j<m_i}$
respectively and they both produce values of type $\tau _{r}$. Additionally, suppose $trans$ is a \textit{recursive morphism} from $\tau_s$ to $\tau_d$.  Under these constraints, the inductive decomposition optimization will eliminate all recursive $trans$ calls and thus, making $trans$ non-recursive.
\end{lemma}

\section{Generalizations of Inductive Decomposition}

Inductive decomposition can be generalized relatively easily to cases when the interpreter takes additional arguments. This is useful, for example, for an interpreter that must take as input the state of the program in addition to an AST. In this case, the specification will have the form 
\begin{formulaarr}
interp_{s}\left(e,~S\right)=interp_{d}\left(trans\left(e\right),~S\right) 
\end{formulaarr}
Here, the inductive decomposition can be modified as follows: 
$$interp_d(\boxed{e'},~ S) \xrightarrow{~~~~~~~~} interp_s(e', ~S)$$

This works because we are replacing what would have been a call to $interp_{d}(\nu, S)$ where $\nu$ would have been the result of calling $trans(e')$ with a call to $interp_s(e', S)$.

A more interesting generalization involves the case when  $trans(e)$  takes some parameters. For example, in cases when the transformation is type directed,  $trans$  may take as a parameter a symbol table which gets updated as part of the recursive calls to  $trans$ . In that case, the specification will look like the one below. 
\begin{formulaarr}
interp_{s}\left(e,~S\right)=interp_{d}\left(trans\left(e, \Gamma\right),~S\right) \\
\mbox{assuming} ~~\Gamma=F(S);
\end{formulaarr}
Without the constraint that $\Gamma$ is a function of $S$, the specification would almost guarantee that $trans$ ignores $\Gamma$, since $\Gamma$ only appears on the right-hand side of the equality.
When $trans$ is a type directed transformation, for example, $F$ would produce a symbol table with the types of the variables in the state $S$. 

In this case, the inductive decomposition algorithm looks as follows:
First, replace $trans(e', \Gamma)$ with the special expression $\boxed{e', \Gamma}$.
Second, apply the following rules  for the evaluation of $\boxed{e', \Gamma}$:

			$\begin{array}{l}
				1. interp_d(\boxed{e', \Gamma}, S) \xrightarrow{~~~~~~~~} interp_s(e', S) \text{ if } \Gamma = F(S)\\
				2. \boxed{e', \Gamma} \text{ in any other case } \xrightarrow{~~~~~~~~} trans(e', \Gamma)
			\end{array}$
            
In this case, the placeholder for $trans$ function must thread its extra parameter $\Gamma$ through its recursive calls, possibly transforming it along the way. 
The transformation of $interp_d$  looks much like it did in the previous generalization, but in addition to adding a call to 
to $interp_s(e', S)$, the optimization will also have to include an check that $\Gamma =F(S)$. 
We found that for simple interpreters and morphisms, this additional check is not a significant constraint; it will hold whenever the state of the transformer is some abstraction of the program state (as is the case with types). The cases where it does not hold, are cases where the state $\Gamma$ tracks some aspect of the program that is not tracked by the interpreter, for example, whether a given construct has been seen or not. In such cases, however, it is relatively easy to extend the interpreter to track this additional aspect as part of its state.

\end{document}